\documentclass{aa}  

\usepackage{graphicx}
\usepackage[varg]{txfonts}
\usepackage{longtable}
\usepackage{natbib}
\bibpunct{(}{)}{;}{a}{}{,} % to follow the A&A style
\defcitealias{Lico2012}{Paper~I}
\begin{document}

   \title{The TeV blazar Markarian 421 at the highest spatial resolution}

   \author{M.\,G.\,Blasi\inst{1,2}\fnmsep\thanks{Email: mariagrazia.blasi@studio.unibo.it}, R. Lico\inst{1,2}, M. Giroletti\inst{1}, M. Orienti\inst{1,2}, G. Giovannini\inst{1,2}, W. Cotton\inst{3}, P.\,G. Edwards\inst{4}, L. Fuhrmann\inst{5}, T.\,P. Krichbaum\inst{5}, Y.\,Y. Kovalev\inst{6,5}, S. Jorstad\inst{7,8}, A. Marscher\inst{7}, M. Kino\inst{9}, D. Paneque\inst{10}, M.\,A. Perez-Torres\inst{11}, B.\,G.~Piner\inst{12}, \and K.\,V. Sokolovsky\inst{6,13}    
           }

   \institute{INAF Istituto di Radioastronomia, via Gobetti 101, 40129 Bologna, Italy
\and Dipartimento di Fisica e Astronomia, Universit\`a di Bologna, via Ranzani 1, 40127 Bologna, Italy
\and National Radio Astronomy Observatory, Charlottesville, 520 Edgemont Road, VA 22903-2475, USA
\and CSIRO Astronomy and Space Science, PO Box 76, Epping NSW 1710, Australia
\and Max-Planck-Institut f\"ur Radioastronomie, Auf dem H\"ugel 69, D-53121 Bonn, Germany
\and Astro Space Center of Lebedev Physical Institute, Profsoyuznaya 84/32, 117997 Moscow, Russia
\and Institute for Astrophysical Research, Boston University, 725 Commonwealth Avenue, Boston, MA 02215, USA
\and Astronomical Institute, St. Petersburg State University, Universitetskij Pr. 28, 198504 St. Petersburg, Russia
\and ISAS/JAXA, 3-1-1 Yoshinodai, Chuo-ku, Sagamihara 252-5210, Japan
\and Max-Planck-Institut f\"ur Physik, F\"ohringer Ring 6, D-80805 M\"unchen, Germany
\and Instituto de Astrof\'{\i}sica de Andalucia, IAA-CSIC, Apdo. 3004, 18080 Granada, Spain
\and Department of Physics and Astronomy, Whittier College, 13406 E. Philadelphia Street, Whittier, CA 90608, USA
\and Sternberg Astronomical Institute, Moscow State University, Universitetskij prosp. 13, 119992 Moscow, Russia
            }

\date{Received ; accepted }

  \abstract
   {High-resolution radio observations allow us to directly image the
     innermost region of Active Galactic Nuclei (AGN). The Very Long
     Baseline Array (VLBA) data analyzed in this paper were obtained during a
     multiwavelength (MWL) campaign, carried out in 2011, from radio to
     Very High Energy (VHE) gamma rays, on the TeV blazar
     Markarian 421 (Mrk 421).}
   {Our aim was to obtain information on the jet structure in Mrk 421
     during the MWL campaign at the highest possible angular
     resolution and with high temporal frequency observations, in
     order to compare structural and flux density evolution with
     higher energy variations.}
   {We consider data obtained with the VLBA at 43 GHz through two sets
     of observations: one is part of a dedicated multi-frequency
     monitoring campaign, in which we observed Mrk 421 once a month from
     January to December 2011 at three frequencies \citep[see][for the
       analysis of the 15 and 24\,GHz data]{Lico2012}; the other
     is extracted from the Boston University monitoring program, which
     observes 34 blazars at 43\, GHz about once per month. We
     model-fit the data in the visibility plane, study the proper motion of jet
     components, the light curve, and the spectral index of the jet
     features. We compare the radio data with optical light curves
     obtained at the Steward Observatory, considering also the optical
     polarization information.}
   {Mrk 421 has a bright nucleus and a one-sided jet
     extending towards the north-west for a few parsecs. The model-fits
     show that brightness distribution is well described using 6--7
     circular Gaussian components, four of which are reliably
     identified at all epochs; all components are effectively 
     stationary except for component D, at $\sim0.4$ mas from the core,
     whose motion is however subluminal. Analysis of the light
     curve shows two different states, with the source being brighter
     and more variable in the first half of 2011 than in the second
     half. The highest flux density is reached in February. A
     comparison with the optical data reveals an increase of the $V$
     magnitude and of the fractional polarization simultaneous with the
     enhancement of the radio activity.}
   {}

  \authorrunning{M.\ G.\ Blasi et al.}

   \maketitle
%
%________________________________________________________________

\section{Introduction}

      \begin{table*}
   \centering
   %\resizebox*{\textwidth}{!}{
   \caption{ Image parameters. The last column lists stations affected
     by significant weather or technical conditions.  }
   \label{tab:table}
   \begin{tabular}{cccccc}
   \hline
   \hline
Observation  & Experiment  & Map peak  & Beam  &  rms    & Notes \\
date in 2011 &    code  &(mJy/beam) &(mas $\times$ mas, $^\circ$)& (mJy/beam)& \\
\hline
Jan 02 & BM303o & 248 & $ 0.42 \times 0.20 $ 15.8 & 1.4 & --- \\
Jan 14 & BG207a & 293 & $ 0.42 \times 0.27 $, 13.7 & 0.23 & MK \\
Feb 05 & BM303p & 245 & $ 0.31 \times 0.19 $, 16.8 & 0.6 & PT \\
Feb 25 & BG207b & 401 & $ 0.47 \times 0.29 $, 20.3 & 0.27 & MK, NL \\
Mar 01 & BM303q & 281 & $ 0.41 \times 0.20 $, $-10.2$ & 0.9 & --- \\
Mar 29 & BG207c & 323 & $ 0.34 \times 0.20 $, $-7.6$ & 0.23 & --- \\
Apr 22 & BM303r & 278 & $ 0.31 \times 0.19 $, $-14.7$ & 0.5 & NL \\
Apr 25 & BG207d & 251 & $ 0.34 \times 0.19 $, $-6.4$ & 0.24 & NL \\
May 22 & BM303s & 317 & $ 0.37 \times 0.17 $, $-13.5$ & 0.8 & BR \\
May 31 & BG207e & 194 & $ 0.35 \times 0.18 $, $-8.1$ & 0.23 & BR \\
Jun 12 & BM303t & 214 & $ 0.33 \times 0.19 $, $-6.5$ & 0.6 & OV \\
Jun 29 & BG207f & 151 & $ 0.34 \times 0.18 $, $-19.1$ & 0.28 & LA \\
Jul 21 & BM303u & 157 & $ 0.36 \times 0.19 $, $-11.0$ & 1.0  & --- \\
Jul 28 & BG207g & 173 & $ 0.40 \times 0.29 $, 9.9 & 0.23  & MK \\
Aug 23 & BM303v & 203 & $ 0.44 \times 0.38 $, $-1.7$ & 0.5 & MK \\
Aug 29 & BG207h & 188 & $ 0.33 \times 0.19 $, $-1.7$ & 0.21 & HN \\
Sep 16 & BM303w & 132 & $ 0.34 \times 0.18 $, $-9.3$ & 0.9 & --- \\
Sep 28 & BG207i & 213 & $ 0.46 \times 0.29 $, 26.9 & 0.28 & MK \\
Oct 16 & BM353a & 214 & $ 0.36 \times 0.22 $, $-13.3$ & 0.7 & --- \\
Oct 29 & BG207j & 207 & $ 0.36 \times 0.20 $, $-9.9$ & 0.21 & --- \\
Nov 28 & BG207k & 260 & $ 0.37 \times 0.20 $, $-10.6$ & 0.20 & FD \\
Dec 02 & BM353b & 179 & $ 0.34 \times 0.18 $, $-11.4$ & 0.7 & --- \\
Dec 23 & BG207l & 248 & $ 0.31 \times 0.17 $, $-9.1$ & 0.38 & FD, HN \\
\hline
\end{tabular} 
\end{table*}

The TeV blazar Markarian 421 (Mrk\,421, $\rm{RA} =
11^{\rm{h}}04^{\rm{m}}27.314^{\rm{s}}$ and $ \rm{Dec} =
+38^{\circ}12'31.80'' $, J2000) is a \textit{high synchrotron peaked}
(HSP) BL Lacertae object at $z=0.031$, making it one of the
closest BL Lacs in the sky.  The spectral energy distribution (SED) of
this object shows two peaks \citep{Abdo2011}: a low-frequency peak due
to synchrotron emission from relativistic electrons in the jet, and a
high-frequency peak probably due to inverse Compton scattering between
the same population of relativistic electrons and the synchrotron
radiation produced by themselves (Synchrotron Self Compton, SSC,
model).
   
Mrk 421 is optically identified with a very bright elliptical galaxy
with a non-thermal and highly polarized component of radiation from
the nuclear region \citep{Ulrich1997}; it also shows high optical
variability \citep{Tosti1998} with two types of behaviour: 
variability on short time scales from
days to hours, and variability on longer time scales of about 23
years.

Mrk421 was the first extragalactic source convincingly detected in
very high energy (VHE, $E > $0.1 TeV) gamma rays \citep[by the 10\,m 
Cherenkov Telescope at the Whipple Observatory,][]{Punch1992}.
It is one of the brightest blazars at X-ray energies, and was one of the few BL Lacs
detected by EGRET \citep{Lin1992}.  As a result, Mrk 421 has
been intensively studied and it has also shown very rapid $ \gamma $-ray
variability on different timescales, from days \citep{Buckley1996}
to $ \sim $ 15 minutes \citep{Gaidos1996}, suggesting that the region
responsible for VHE emission must be very small. This has led to the
expectation of extreme values of the Doppler factor, but radio data
analysed in previous works did not provide evidence for superluminal
motion of jet components \citep{Piner1999,Piner2010,Giroletti2006}.
   
In this context, and after a first large multi-wavelength campaign
organized in the {\it Fermi} era \citep{Abdo2011}, we set up a VLBA
multi-frequency campaign to observe Mrk421 at 15, 24, and 43\,GHz, in
full polarization, once per month during 2011, to complement efforts
carried out at various observatories at higher energies. In
\citet[][hereafter Paper~I]{Lico2012}, we reported an analysis of the
15 and 24\,GHz data.
The results of this work show that at these frequencies the structure of Mrk 421 results 
dominated by a compact ($ \sim $0.13 mas) and bright nucleus, 
characterized by a flat spectral index ($ \alpha \sim -0.3 \pm 0.2$), 
with a one-sided jet detected out
to $ \sim $10 mas, that presents steeper spectral index ($ \alpha \sim -1.2 \pm0.5 $).
Model-fitting of the source is well described by 5-6 components that are consistent with being stationary 
during this period, while the flux density analysis shows significant variations for the core, which is brighter during the first part of 2011. The observational properties suggest that the jet bulk velocity 
could be different between the radio and the high-energy emission regions: under a small viewing angle ($ 2^{\circ} < \theta < 5^{\circ} $), the inferred high energy and radio Doppler factors are respectively and $ \delta_{\rm{h.e.}} \sim$ 14 and 
$ \delta_{\rm{r}} \sim $ 3).

In this paper, we focus on the 43\,GHz data, which
of all astronomical bands provide the best angular resolution
achievable with a regular cadence. In addition, we consider data from
the Boston University blazar monitoring project to sample the
evolution of the radio jet with even higher cadence. We note that
several more detailed publications are in preparation about the results of the
multiwavelengthm (MWL) campaign; here we provide a first simple comparison
between high resolution radio observations and optical data.

\citet{Wang2004} estimated the mass of the central supermassive
black-hole in Mrk\,421 using the relation between the mass and the
velocity dispersion, obtained by the broadening of $ \rm{H}_{\alpha} $
line, and they deduced $ M_{\rm{BH}}= 10^{8.29} M_{\odot}$.
Considering that our VLBA images achieve an angular resolution of
$\sim 0.2$\,mas, in this paper we will focus on structures of the order
of about $ 2\times 10^{4} $ Schwarzschild radii.

The paper is organized as follows: in Sect.\ 2 we describe the data used
for model-fitting and for the MWL study; in Sect.\ 3 we present our
results obtained by model-fitting, flux density analysis, and the MWL study. 
In Sect.\ 4 we discuss the astrophysical implications of the results and
we present our conclusions in Sect.\ 5.

We adopt the following cosmological parameters for a flat universe: $
h_0=0.73\, \rm{km \, s^{-1} \, Mpc^{-1}}$, $ \Omega_{\rm{m}}= 0.27$, $
\Omega_{\Lambda}= 0.73$ \citep{Spergel2003}.  Then for $z=0.031$, we obtain a linear scale
of 0.593\,pc/mas. The spectral index $\alpha$ is defined such that $
S(\nu) \propto \nu^{-\alpha}$.

\section{Observations}

\subsection{Radio data}

   \begin{figure*}
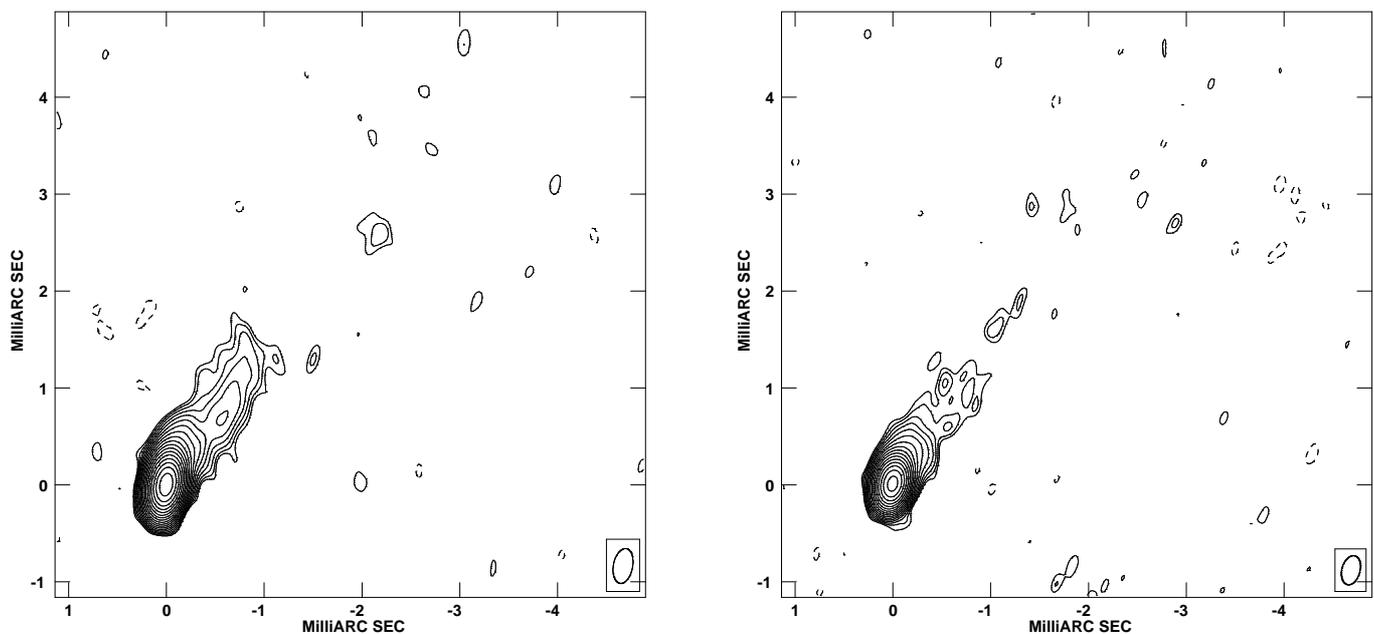

	\centering
	\includegraphics[scale=0.46]{fig1a.eps} 
	\qquad
	\includegraphics[scale=0.46]{fig1b.eps} 
    \caption{(Left) VLBA image of Mrk 421 at 43\,GHz corresponding to the
      November observation of BG207 campaign, contoured at $(-1, 1,
      \sqrt{2}, 2, \ldots) \times 0.55$ mJy/beam.  (Right) VLBA image
      of Mrk 421 at 43\,GHz corresponding to April observation of the
      Boston University monitoring program, contoured at $(-1, 1,
      \sqrt{2}, 2, \ldots) \times 1.3$ mJy/beam. \label{fig:mappe}}
   \end{figure*}

In this paper, we consider two datasets obtained with the VLBA at 43\,GHz.  
The main one consists of 12 observations obtained once a month
throughout 2011.  At each epoch the source was observed for a net
observing time of little more than 4 hours, distributed evenly over 8 hours.
We observed three calibration
sources: J0927+3902, J1310+3220, and J0854+2006 that we also used to
check the amplitude calibration.  The output signal from the
correlator was divided into four IF with bandwidth of 16\,MHz, each 
split into 32 channels with separations of 500\,kHz.  Phase and
amplitude calibration of the fringe visibilities has been done with the
AIPS \footnote{http://www.aips.nrao.edu/index.shtml} software package, while self-calibration procedures and imaging
has been performed with Difmap Shepherd et al. 1994, 1995) software package.
   
In addition to the main dataset, we also include other VLBA
observations at 43\,GHz of Mrk 421 for a more detailed study of the
source structure.  These data (11 epochs in total) were provided by
the Boston University blazar group\footnote{\texttt{http://www.bu.edu/blazars/VLBAproject.html}  }
and belong to their monitoring
program of gamma-ray blazars, which observes 34 blazars at 43\,GHz
about every once per month with the VLBA.  These observations are
characterized by shorter time duration: the total observation time for
the source is about 40 minutes for each epoch, so the $ (u, v) $-plane
coverage is more sparse and the sensitivity lower than in the main 
43\,GHz dataset.
   
Table \ref{tab:table} reports some useful information such as the size
of the beam, the final rms noise, and antennas affected by significant
failures at each observation.  The two experiment codes, BG207 and
BM303/BM353, denote that observations belong to the
broadband campaign or to the Boston University program, respectively.
      
\subsection{Optical data}

For the MWL comparison, we consider optical data obtained at the
Steward Observatory of Arizona University\footnote{{\tt
    http://james.as.arizona.edu/\~{}psmith/Fermi/\#mark2}}.  The
Steward Observatory monitors the optical linear polarization of a
blazar sample using the SPOL CCD Imaging/Spectropolarimeter, and it
yields also measurements of the brightness and spectral index of the
optical synchrotron light \citep[see][for an overview of the optical
  program]{Smith2009}.

For Mrk 421, we consider observations between 2011 Jan 2 (MJD 55563)
and 2011 Dec 31 (MJD 55926), characterized by the uneven coverage
typical of optical observations. In total, we have useful data for $V$
magnitude (41 data points), and information about polarization
percentage and position angle (49 points).

\section{Results}

\subsection{Images}

Figure \ref{fig:mappe} shows two sample images of the source, one from
the main dataset and one from the Boston University data (in the left
and right panel, respectively). We obtained the clean images with
Difmap and produced the final contour plots with AIPS.  The maps
of Mrk 421 reveal a bright and variable core:\ the mean peak
brightness is $\langle B \rangle = 233 $ mJy beam$^{-1}$ at 43\,GHz,
and the standard deviation is $ \sigma_B = 60$ mJy beam$^{-1}$,
suggesting significant scatter in the core brightness.  A collimated
one-sided jet extends on a scale of few parsecs towards the north-west;
the \textit{position angle} (PA) of the jet is about $ -35^{\circ} $
for $ \sim 4$\,mas ($ \sim 2.4$\,pc).
 
Lower frequency maps of Mrk421 from the literature
\citep[e.g.][]{Giroletti2006,Piner1999,Giovannini1999}, show a bright
and collimated jet extending up to about 20\,pc, which then widens and
becomes fainter, also showing bending and diffuse emission on a scale
of about 100\,mas. Our higher frequency, hence higher resolution,
images provide a better view of the innermost region, the closest one
to the central black hole. We obtain a better estimate of the core
size and flux density, and reveal finer details of the transverse
structure, such as the brightness asymmetry at $\sim$1\,mas from the
core, where the western limb is slightly brighter than the eastern
one.  Such details are in agreement and lend support to the finding of
previous, lower sensitivity, images of the source at the same
frequency \citep{Piner1999,Piner2010}.

\subsection{Model-fit}

   \begin{figure}
	\centering
	\includegraphics[scale=0.46]{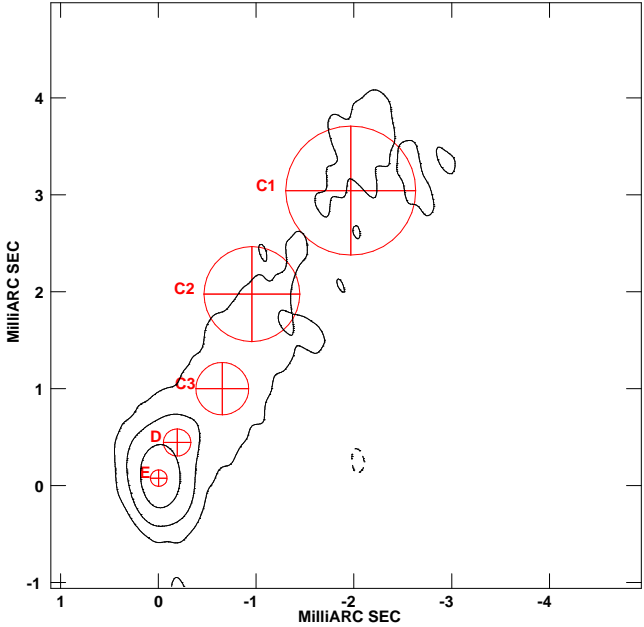} 
    \caption{Modelfit components (in red) from the 2011 July 28 epoch overlaid with contours from a stacked image of all epochs. The circular Gaussian components are shown, while the reference delta is not plotted for clarity reason. Contours are traced at $(-1, 1, 10, 100) \times 0.4$ mJy/beam. \label{f.modelfit}}
   \end{figure}

The \textit{model-fit} technique, executed with Difmap, allows us to
describe the brightness distribution of the source using a few simple
components.  Analyzing the components found from epoch to epoch we can
determine their apparent proper motion. We consider both our main
dataset and the additional 43\,GHz VLBA data, provided by Boston
University, in order to have more constraints for the identification of
components from epoch to epoch.  However, the latter observations are
characterized by short time durations, so the model-fit procedure is
only sensitive to the main features of the source.
   
For both datasets we used a point source (delta function) to describe
the central region of the core, which we assume to be stationary, while
the jet was fitted with circular Gaussian components. We show in
Table~\ref{tab:dati_model-fit} the parameters resulting from the
model-fit procedure, epoch by epoch:\ the total flux density $S_{\rm
  tot}$, the distance from the core in polar coordinates $r$ and
$\theta$, and the component size $a$.  We estimate flux density
uncertainties considering both a calibration error of about 10\% and a
statistical error given by the rms of the map.  Position errors were
estimated according to the formula:\ $ \Delta r= a/(S_{\rm peak}/{\rm
  rms}) $ where $S_{\rm peak}$ was calculated considering that each
component is a circular Gaussian component, finally getting: $ S_{\rm
  peak}=(2.355)^{2} \times (S_{\rm tot}/2\pi a^{2}) $.  For each epoch
there are some very bright components smaller than the beam size (they
are not resolved), whose nominal uncertainties as a result are too small, so
the error value is set equal to 10\% of the beam size.  There are also
some very faint extended components, whose S/N ratio is less than 3.
In this case the uncertainty is calculated by the formula $ \Delta r=
a/(S_{\rm tot} / \sqrt{n} \, {\rm rms}) $, where $ n $ is the number of times
that the beam is included in the full width at half maximum (FWHM) of the component.
   
Overall, we obtain a good description of Mrk 421 with 6 or 7
components; comparing model-fit results of both datasets we note a
good agreement with exception of the farthest component. This is not
surprising, given the shorter duration of the runs in the BM303/BM353
series.  In particular, by analysing the position with respect to the
core and the flux density of each component from epoch to epoch, we
identified across epochs five components in addition to the
core (see Fig.~\ref{f.modelfit}).  Comparing our data with the 15 and 24\,GHz datasets, analysed in
\citetalias{Lico2012}, we find a good agreement in the outer part of
the jet, at $r>$1 mas from the core, both in position and flux density
(taking into account the spectral index of the components).  Therefore
we label the components in this region in the same way, i.e., with C1,
C2 and C3.  We note however that component C2 is barely detected at 
43\,GHz, consistent with its low flux density (a few mJy) and steep
spectrum ($ \alpha =-1.2 \pm 0.5 $) at 15 and 24\,GHz
\citepalias[see][]{Lico2012}.  Components within 1\,mas of the core
are difficult to interpret and identify, even if we compare them with
components identified in \citetalias{Lico2012}. Indeed, the spatial
frequencies accessible at 43\,GHz allow us to describe the brightness
distribution of the innermost region by a larger number of Gaussian
components than the lower resolution data.  We thus introduce two new
components, labelled D and E, the first of which is identified only
from February to August and it is usually placed between 0.3 and 
0.5\,mas from the core, while the second is the closest to the core 
($<$0.1\,mas).
   
Analysing the parameters obtained for the components, it is
interesting to note that components farther than 0.4\,mas from the core
moslty have PAs between $ -20^{\circ} $ and $ -40^{\circ} $, while the
closest components have quite variable PAs. In general, this is not too
surprising, since the polar angle is more variable at small
radii. However, the behaviour of component E is particularly
remarkable: its PA is systematically off the main jet axis,
positioning mainly to the north-east with respect to the core
(PA$>0^\circ$). This suggests that there is a bending of the jet in
the region closest to the core ($<$0.5\,mas from the core), with the
jet emerging towards the north-east and then changing direction to the
north-west.  This characteristic was also observed in the same
component by \citet{Piner1999,Piner2010}, as well as in other BL Lac
objects on similar angular scales, \citep[e.g.\ in Mrk\,501,]
[]{Giroletti2004}. We further note that Mrk\,421 itself shows
several PA changes out to the kiloparsec scale radio emission
\citep{Giroletti2006}.   
   
\subsection{Apparent speed}

The identification of the components from epoch to epoch is important
for studying their motion with respect to the core, which is considered
to be the stationary reference point of our frame.  We have
calculated the apparent speed of these components during 2011 making a
linear fit of the separation from the core of identified components
over the observation time range.  For C1 we have fitted only our BG207
campaign VLBA data, while for C3, D and E we used both our VLBA data
and the Boston University observations.  Table~\ref{tab:dati moto} shows
results of linear fits to the data.
   
   \begin{table}
   \centering
   \caption{
   Apparent speed of identified components resulting from linear fits.}
   \label{tab:dati moto}
   \begin{tabular}{cccc}
   \cline{1-3}
   \hline
   \hline
   \multicolumn{4}{c}{Apparent speed} \\
   \hline
   Component & mas/yr & $ \beta_{\rm app} $ &  $ \beta_{\rm app,\ Paper I} $ \\
   C1 &  $ 0.12 \pm 0.13 $  &  $ 0.24 \pm 0.25 $ & $ 0.34 \pm 0.24$  \\
   C3\tablefootmark{\textit{a}} & $ -0.11 \pm 0.04 $ & $ -0.20 \pm 0.07 $ & $\ldots$\\
   C3\tablefootmark{\textit{b}} & $ 0.02 \pm 0.04 $ & $ 0.04 \pm 0.08 $ & $ 0.10 \pm 0.11$\\
   D & $ 0.24 \pm 0.04 $ & $ 0.46 \pm 0.08 $ & $\ldots$\\
   E & $ 0.001 \pm 0.019 $ & $ 0.003 \pm 0.038 $ & $\ldots$\\
   \hline
   \end{tabular}
   \tablefoot{
   \tablefootmark{(\textit{a})}{Linear fit for C3 dataset including C3* (see Table \ref{tab:dati_model-fit}) data.}
   \tablefootmark{(\textit{b})}{Linear fit for C3 dataset excluding C3* (see Table \ref{tab:dati_model-fit}) data.}
   }
   \end{table}
   
\begin{figure}
\centering
\includegraphics[width=\columnwidth]{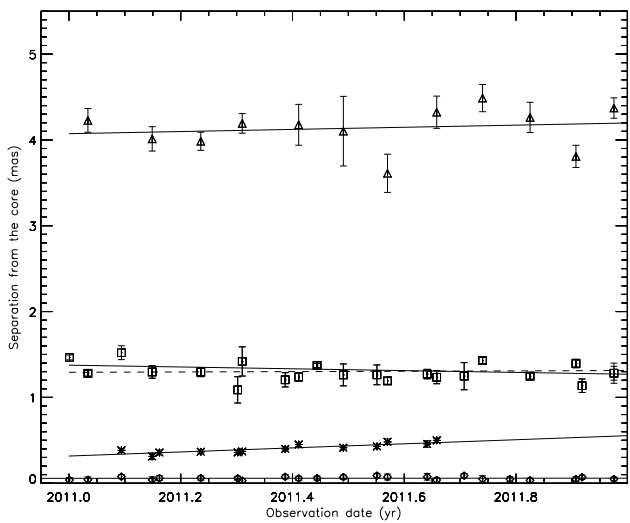}
\caption{Linear fit for the component C1, C3, D and E.  For C3 we have
  two datasets: $\rm{C3}^{\textit{a}}$ (solid line) and
  $\rm{C3}^{\textit{b}}$ (dashed line). \label{fig:moto_componenti}}
\end{figure}

The resulting apparent speeds suggest that components C1 and E do not have
significant proper motions --- they are consistent with being stationary
and their apparent speeds would be at most a few times $0.1\, c$.
Component D has a subluminal proper motion at the 5.8$ \sigma $ level
($\beta=v/c= 0.24 \pm 0.04 $).  We consider two scenarios for C3,
using two different datasets. 
This is necessary because the identification of C3 in the
January and February data of the Boston University observations (indicated
by C3* in Table \ref{tab:dati_model-fit}) is less secure, since its
separation from the core is not consistent with the one measured in
the BG207 series in the same months.  The results obtained through the two
linear fits are indicated in Table \ref{tab:dati moto} with superscripts
\textit{a} and \textit{b}.  The first linear fit, indicated by
C$3^{\textit{a}}$ in Table \ref{tab:dati moto}, includes the January and
February data from the Boston University observations; in this case we see
that C3 seems to move towards the core with a subluminal speed with a
significance of about 3$\sigma$.  The second linear fit
(C$3^{\textit{b}}$) excludes C3* data and we obtain an apparent speed
of $ \beta = 0.04 \pm 0.08 $, which is consistent with no motion.

\subsection{Radio flux density analysis}

Analysing the trend of radio flux density at the various epochs, it
emerges that the more external components, C1, C2, C3 and D are the
faintest components, and they do not show significant variations. A few
outliers occur for each of them but these are more likely to be
statistical fluctuations due to model-fit artefacts on faint and
extended features.

\begin{figure}
\centering
\includegraphics[width=\columnwidth]{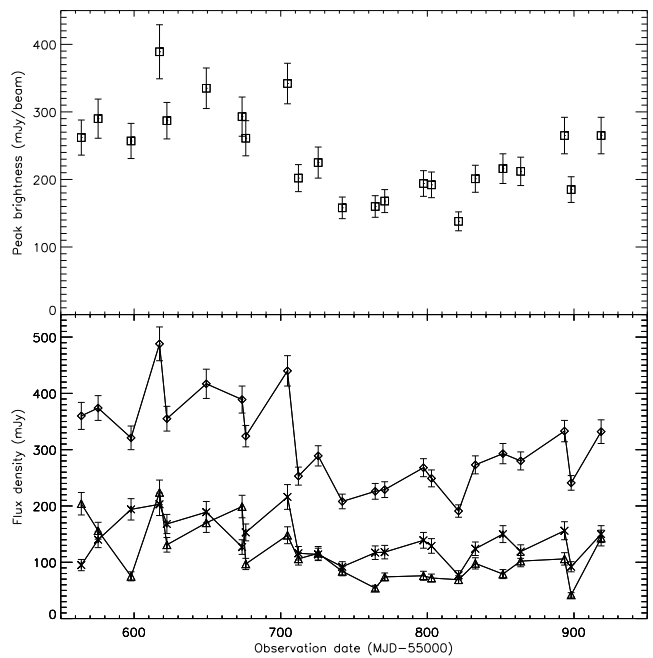}
\caption{(Top panel) Map peak brightness light curve from images
  restored with a 0.3\,mas beam.  (Bottom panel) Flux density of the
  map peak (diamonds), core (asterisks) and component~E (triangles)
  during 2011.\label{fig:flux multi}}
\end{figure}

On the other hand, the model-fit results on the most compact
components, and the image peak itself, suggest that the core is
genuinely variable.  Thus, for the sake of uniformity, we convolved
the 23 images with a circular beam with a radius of 0.3\,mas, which is a
conservative representation of the beam for VLBA data at 43\,GHz.
After this procedure we analyse the trend of map peak brightness
through the 23 epochs, that we show in Figure~\ref{fig:flux multi}
(top panel).  Although there is a large scatter from epoch to epoch,
we clearly see two different behaviours throughout 2011.
During the first part of the year, the peak goes through a phase of
variability and enhanced brightness; it increases and reaches a peak
value in February (MJD 55617), then it decreases until July.  In the
second part of the year, from July to December, the overall behaviour
of the peak brightness shows a lower level of activity and a more
regular trend, with a slow increase of the brightness.
   
The bottom panel in Figure~\ref{fig:flux multi} shows that the light
curves for the core and component~E largely overlap and intersect each
other. These two components are the brightest features in Mrk\,421 and
they are responsible for most of the total flux density of the source;
since most of the flux density is collected within a radius of about
0.3\,mas from the core, we confirm that Mrk\,421 is a strongly
core-dominated source.  The overlap between the two light curves also
suggests that the core and component~E (whose separation is about one
beam size) are really the same, slightly extended feature undergoing
a single dissipation process.   
   
\subsection{Multiwavelength observations}

At higher energies, Mrk\,421 is a highly variable blazar; it has shown
different episodes of dramatic flares since it was observed, among
which one of the brightest X-ray and VHE flares ever seen in February
2010 \citep{Isobe2010,Shukla2012} and an even brighter, recent one in
March/April 2013 \citep[e.g.][]{Cortina2013,Paneque2013}.  Comparing
multiwavelength observations is interesting for studying the origin of
the radiation and the size of emission region and eventually understanding
not only the mechanisms that lead to the jet formation, but also the
relation between jets, central black hole and accretion disk.

Indeed, the radio data presented in this paper and in
\citetalias{Lico2012} are obtained in the framework of a large
project, which involves multi-epoch and multi-instrument observations
at different bands: in the sub-mm (SMA), optical/IR (GASP), UV/X-ray
(Swift, RXTE, MAXI), $\gamma$-rays (Fermi-LAT, MAGIC, VERITAS), and
at cm wavelengths (e.g., F-GAMMA, Medicina). While a series of
dedicated papers are in preparation, we report here some first basic
findings based on the correlated analysis of our radio data and the
optical data from the Steward Observatory.
  
As we saw previously, 43\,GHz emission during 2011 shows a peak in
February observation (MJD 55617) and variability throughout the first
half of the year: data show that the flare seems to have a duration of
about $\Delta \tau \sim 150$ days (from MJD $\sim 55563$ to MJD $\sim
55712$), corresponding to an emission region of about $ 3.9 \times
10^{17} \mathrm{cm}$, or 0.13~pc, considering the causality argument
which implies the size of emission region must be $ R < c \Delta
\tau $.  The flare could be somewhat longer since we do not consider
data before 2011 January 2. However, during the second part of 2011
Mrk\,421 does not show any flares; the total flux density has a minimum
in July (MJD 55764), then it slightly increases until the end of 2011.
   
\begin{figure}
\centering
\includegraphics[width=\columnwidth]{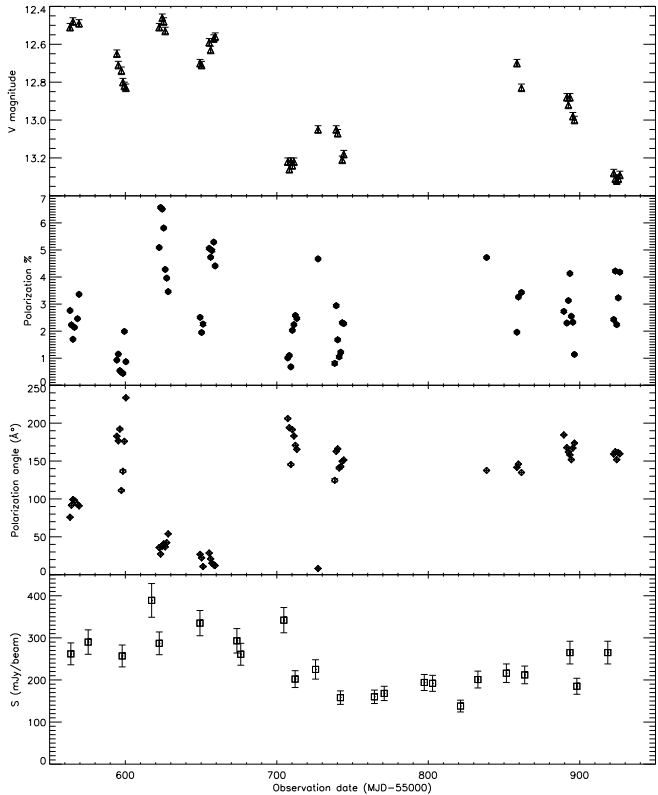}
\caption{V magnitude, optical polarization percentage and angle, and
  map peak brightness of Mrk\,421 during 2011 (from MJD 55550 to MJD
  55950).\label{fig:MWL}}
\end{figure}
   
In Figure \ref{fig:MWL} we show optical data (V magnitude,
polarization percentage and angle) and radio brightness of the map
peak during 2011.  Comparing radio to optical data we see that radio
peak is almost simultaneous with an optical peak (MJD 55624).  It is
interesting to note that just before the enhanced activity, from MJD
55594 to MJD 55600, the V magnitude decreases, the associated polarization
angle has a wide scatter of about $ 150^{\circ}$, and the polarization
percentage is very low ($\sim$1--2\%); while during the flare the
polarization percentage reaches values of about 7\% and the polarization
angle has a very small scatter, of about $ 20^{\circ}$.

It is also interesting to note that the public high energy data by {\it 
Fermi}\footnote{\texttt{http://fermi.gsfc.nasa.gov/ssc/data/access/lat/msl\_lc/source/Mrk\_421}} 
show a similar behaviour during 2011 when compared with our  data.
During the period between MJD 55550 and MJD 55950, the $ \gamma $-ray light curve reveals 
an enhanced activity in the first part of the year, in agreement with our results 
in the radio and optical bands. However, we remind that a dedicated $\gamma$-ray analysis 
and much more detailed MWL study are in preparation.

\section{Discussion}

\subsection{Comparison with 15 and 24 GHz data}

We compare our 43\,GHz results with the 15 and 24\,GHz data, analyzed in
\citetalias{Lico2012}, belonging to the same monitoring campaign.
Images confirm the same structure on the parsec-scale: a dominant central
compact nucleus, from which a one-sided jet extends towards
north-west, with a PA of about $ -35^{\circ} $.
   
The difference is angular resolution, which increases going to higher
frequencies, as we can see from the beam size: for January
observation we obtain $ 0.89 \, \rm{mas} \times 0.52 $ mas at 15\,GHz,
$ 0.73 \, \rm{mas} \times 0.41 $ mas at 24\,GHz and $ 0.42 \, \rm{mas}
\times 0.27 $ mas at 43\,GHz.  This is also clearly visible in the 
model-fit results.  At 15\,GHz we need five Gaussian components, while
the 24\,GHz data are well described by six Gaussian components. In
\citetalias{Lico2012}, we pointed out that the second innermost 15\,GHz
component splits into two components at 24\,GHz because of the higher
angular resolution.  At 43\,GHz the brightness distribution is well
described by six or seven components: we further resolve and better
constrain the inner features, while we become less sensitive to the
more extended regions, like C2, which are also depressed due to their
steep spectrum nature.

We also find good agreement between the 43 and the 15 and 24\,GHz datasets
in terms of apparent motions, with linear fits to 15, 24, and 43\,GHz 
data always yielding low values for apparent speeds.  In
particular, our findings are fully consistent with those of
\citetalias{Lico2012} for component C1 and a good match is also found
for C3 if we consider our C3$^{\textit{b}} $ result.
Moreover, our results are also in agreement with the most recent 
MOJAVE 15 GHz VLBA measurements  \citep{Lister2013}.
   
Analysing flux densities we note similar light curves at all three
radio-frequencies, with highest values in the first half of 2011; the
maximum is reached somewhat earlier at 43\,GHz, with a prominent peak
in February, while at the lowest frequencies the values of February
and March are quite similar. The overall variability also increases
with frequency, as seen by the values of the modulation index
\citep[e.g.][]{Richards2011}:

$$
m_{\rm data} = \frac{\sqrt{\frac{1}{N}\sum_{i=1}^N\left(S_i- \frac{1}{N}\sum_{i=1}^NS_i
\right)^2}}{\frac{1}{N}\sum_{i=1}^NS_i}\,.
$$

Indeed, the modulation index calculated for the image peak rises from
$m_{15}=0.18$ to $m_{43}=0.27$. The effect would become even larger if
the total flux density were considered, since the extended and less 
variable components are more prominent at lower frequencies. This is
clearly visible from the spectral index value. We estimate an average
spectral index both for the innermost region ($ <$ 1 mas from the
core) and for the external region ($>$ 1 mas from the core), by
fitting a power law $\nu^{-\alpha}$ to the mean flux density at three
radio frequencies.  We find $ \alpha = 0.10 \pm 0.29 $ for the
innermost region, and $ \alpha = 0.64 \pm 0.27 $ for the outermost
region. The relatively large uncertainties derive from the somewhat
crude assumption of a power law, since some spectral curvature is
present, in particular in the core.   
   
\subsection{Proper motion and Doppler factor}

Our results lend further support to the growing evidence of lack of
superluminal motions in VHE blazars in general, and in Mrk\,421 in
particular. Besides the present work and \citetalias{Lico2012}, other
studies on Mrk 421 have shown jet components with subluminal speeds
\citep{Piner1999,Piner2005,Piner2010}. A recent work has reported
inward radial motion for one component soon after the large X-ray
flare of Mrk\,421 occurred in 2010 mid-February \citep{Niinuma2012}. In
the latter work, the authors propose that the inward component motion
is actually due to a shift of the centroid of the core component
following the ejection of an unresolved superluminal component. This
is not the case for our component C3. Therefore, it could indicate
that superluminal components, if any, tend to be detected soon after
giant flares rather than in quiescent states. On the other hand, we
can calculate the distribution of the linear velocity estimated from
any set of four consecutive measurements of the C3 position in our
campaign. It turns out that 5/18 sets ($\sim28\%$) yield a negative
velocity as large as that reported by \citet{Niinuma2012}. Although
this does not rule out the previous hypothesis, it nonetheless
highlights how components wander around the line of best fit, and that
it can be risky to extrapolate apparent short-term motions too far.

In any case, the lack of superluminal motions on $\sim 1$-yr timescale
for all components remains somewhat surprising for blazars which
showing high variability at TeV energies. Such variability and the presence of
VHE emission itself require quite a high Doppler factor, $\delta$, to
avoid suppression of $\gamma$-ray emission region through pair
production. Various works show how $\delta \geq 10$ is naturally
required in quiescent states, and even higher values are needed during
rapid variability events
\citep{Gaidos1996,Albert2007,Donnarumma2009,Abdo2011}. In the radio
band, Mrk\,421 shows a one-sided jet structure and a large core
dominance value, which, as widely discussed in the literature, imply
it is subject to mild relativistic \textit{boosting} effects, with
$\delta, \Gamma \sim$ a few. Thus, even though the jet is likely to be
relativistically boosted in both bands, completely different levels of
boosting are typically required. We refer to \citetalias{Lico2012} for
a detailed discussion of the proper motion, jet velocity, viewing
angle, and Doppler factor estimate (Sects. 3.5, 3.6, and 4) and
briefly highlight below the main implications of the results from our
campaign.

\begin{itemize}

\item The consistent values found at 15, 24, and 43\,GHz for the polar
  coordinates and the sensible spectral behaviour of each jet
  component suggests that they provide a well defined description of
  the jet structure; however, they are much less prominent features
  than the typical bright, compact knots found in flat spectrum radio
  quasars (FSRQ).

\item None of the components in Mrk\,421 shows apparent superluminal proper
  motion, again in sharp contrast to the behaviour of many FSRQs and
  apparently at odds with various constraints on the Doppler factor in
  Mrk\,421 itself.

\item Statistical arguments rule out the possibility to explain such
  oddity by invoking an extremely small viewing angle
  \citep[e.g.][]{Lister1999,Tavecchio2001}; ultimately, we have to
  conclude that the \textit{pattern velocity} is not a good representation
  of Mrk\,421's jet \textit{bulk velocity}.

\item The fact that the pattern speed is generally not a good
  representation of the bulk speed is generally accepted also for more
  powerful blazars; however, in those sources components moving at
  close to the bulk speed are detected occasionally if long enough
  monitoring projects are carried out; the fact that in over 17 years
  of VLBI monitoring not one single component has been seen in Mrk\,421
  travelling at very large velocity suggests that the plasma is
  probably not moving at highly relativistic bulk speeds in the radio
  jet.

\item The jet bulk velocity structure is complex, with different
  values for the gamma-ray and the radio emitting regions, either
  along or across the jet; in the latter case, we could have a
  \textit{spine-layer} model, with the jet transversally structured
  with respect to its axis and consisting of an inner spine,
  characterized by higher velocity and the outer layer, which
  decelerates because of interaction with external medium, proceeding
  more slowly.

\item In support of the presence of a velocity structure across the
  jet axis, our images show evidence of a transversally resolved
  brightness structure (see
  Fig.~\ref{fig:mappe}). \citet{Giroletti2006} suggest a
  limb-brightening structure in Mrk\,421 at distances down to 2\,mas
  from the core, while \citet{Piner2010} argue evidence of
  limb-brightening structure on smaller scale, at least 0.25\,mas from
  the core.
   
\item All the evidence for complex velocity structures in the VLBI jet
  of Mrk\,421 and other sources implies that many of the estimates of
  jet emission parameters from high-energy SSC modeling that used
  one-zone homogeneous sphere models are likely to be
  over-simplifications.

\end{itemize}

\subsection{Variability and multi-frequency analysis}

Whereas Mrk\,421 has shown dramatic variability in the X-ray and VHE
bands, it has traditionally been quite stable at radio
frequencies. Therefore, our finding of an enhanced variability in the
first half of 2011 is quite remarkable. Even more, the source has then
shown a dramatic radio flare at 15\,GHz in September 2012, with an
increase of approximately 2.5 times well fit by an exponential curve
with a doubling time of 9 days \citep{Hovatta2012}. No new jet knots
have appeared in our VLBA images after the February 2011 peak, showing
that the cooling is very efficient and takes place within the 43\,GHz
beam.

A further interesting piece of information comes from the comparison of the
radio and optical data (see Fig.~\ref{fig:MWL}) and in particular of
the optical polarization. We observe an increase of the polarization
fraction during the radio and optical flare, associated to a
remarkably small ($\sim 20^\circ$) scatter of the electric vector
position angle (EVPA). This suggests that the flare is associated with
the transition from a chaotic and turbulent magnetic field in the
emission region to a more ordered state, that is responsible for
optical and radio luminosity increase.  The near simultaneity of the
optical and radio flares indicates that optical and radio emission
region are co-spatial and given the light curves of the radio
components we can constrain it to be within $\sim 0.2\ \rm{mas} \sim
3.7\times 10^{17}$\,cm.

Later in the year, the optical emission continues to be variable but
it never reaches again the same value of optical magnitude ($V=12.46$)
and percentage polarization ($p=0.066$). The discussion of the optical
and MWL behaviour of the source in 2011 will be the subject of future
papers.

\section{Conclusions}

The new results presented and discussed in this paper provide both 
strong support of previous discoveries on TeV blazar jets as well as
some new interesting findings. We confirm the presence of a
limb-brightened jet structure near the radio core region, the lack of
a high velocity proper motion, and the absence of bright substructures
(knots) in the jet. We note that in FSRQs fast proper motions have been
measured in bright knots or substructures present in radio
jets. The uniform brightness of the Mrk 421 jet not only prevents a
reliable proper motion measure but suggests a physical difference
between jets in FSRQ and BL Lac sources.

In addition, thanks to the high observing frequency and the dense time
sampling, we have a sharp view on the inner core which provides
evidence of significant and somewhat unexpected variability in the
core region. Such variability presents an intriguing correlation with
the optical behaviour of the source in optical magnitude and
percentage polarization, hinting at a significant evolution of the
magnetic field playing an important role in the emission process.

We anticipate that further insight into the physics of Mrk\,421
and of TeV blazars in general will be obtained from the analysis of
the VLBA polarimetry data and from the dedicated time and spectral
study of the whole broadband (radio to gamma rays) 2011
campaign. These works will also provide a reference for the analysis
of MWL data from recent and unprecedented flaring episodes across the
whole electromagnetic spectrum
\citep{Hovatta2012,Cortina2013,Paneque2013}.

\begin{acknowledgements}

   This work is based on observations obtained through the BG207, BM303, and BM353 VLBA
   projects, which make use of the Swinburne University of Technology
   software correlator, developed as part of the Australian Major
   National Research Facilities Programme and operated under licence
   (Deller et al.\ 2011). The National Radio Astronomy Observatory is
   a facility of the National Science Foundation operated under
   cooperative agreement by Associated Universities, Inc. We
   acknowledge financial contribution from grant PRIN-INAF-2011. This
   research is partially supported by KAKENHI (24540240). K.V.S.\ and
   Y.Y.K.\ are partly supported by the Russian Foundation for Basic
   Research (project 11-02- 00368 and 12-02-33101), 
   and the basic research program
   ``Active processes in galactic and extragalactic objects'' of the
   Physical Sciences Division of the Russian Academy of
   Sciences. Y.Y.K.\ is also supported by the Dynasty Foundation. The
   research at Boston University was supported in part by NASA through
   Fermi grants NNX08AV65G, NNX08AV61G, NNX09AT99G, NNX09AU10G, and
   NNX11AQ03G, and by US National Science Foundation grant
   AST-0907893.  This study makes use of 43\,GHz VLBA data from the
   Boston University gamma-ray blazar monitoring program
   (http://www.bu.edu/blazars/VLBAproject.html), funded by NASA
   through the Fermi Guest Investigator Program.  Data from the
   Steward Observatory spectropolarimetric monitoring project were
   used. This program is supported by Fermi Guest Investigator grants
   NNX08AW56G, NNX09AU10G, and NNX12AO93G.

\end{acknowledgements}

\clearpage
\onecolumn
\longtab{3}{
\begin{longtable}{lcccccccccc}
\caption{Model-fit parameters for each component.}\label{tab:dati_model-fit} \\
\hline
\hline
Epoch & experiment & Component & $S_{\rm tot}$ & $\Delta S_{\rm tot}$ &  $r$  & $\Delta r$ & $\theta$ &  $a$  \\
         &   code     &            &   (mJy)     &    (mJy)     & (mas) &   (mas)    & ($^\circ$) & (mas) \\
\hline
\endfirsthead
\caption{continued.} \\
\hline
\hline
Epoch    & experiment & Component  & $S_{tot}$ & $\Delta S_{tot}$ &   r   & $\Delta r$ &  PA   &  a  \\
         &   code     &            &   mJy     &    mJy           & (mas) &   (mas)    & (deg) & (mas) \\
\hline
\endhead
\hline
\endfoot
2011 Jan 02 & BM303o &  Core  &   95   &   10  &   Ref  &   0    &    Ref   &  0  \\*
            &        &   E    &  204   &   20  &   0.04 &   0.03 &     0.3  &  0.15  \\*
            &        &   --   &  25.5  &    3  &   0.49 &   0.05 & $-$47.1  &  0.42  \\*
            &        &   C3*  &   8.5  &   1.6 &   1.47 &   0.03 & $-$41.5  &  0.14  \\*
            &        &   C2   &  10.3  &   1.7 &   2.01 &   0.03 & $-$26.3  &  0.16  \\*
            &        &   --   &  16.2  &   6.4 &   3.4  &   0.5  & $-$30.1  &  1.28  \\        
         
2011 Jan 14 & BG207a &  Core  &  140   &  14   &   Ref  &  0     &   Ref    &  0   \\
            &        &   E    &  156   &  15   &   0.04 &  0.03  &    81.7  &  0.11 \\
            &        &   --   &   42   &   4   &   0.16 &  0.03  & $-$22.3  &  0.30 \\
            &        &   --   &   12.8 &   1.3 &   0.64 &  0.03  & $-$36.9  &  0.34 \\
            &        &   C3   &   12.8 &   1.3 &   1.28 &  0.03  & $-$37.9  &  0.54 \\
            &        &   C1   &    9.8 &   1.4 &   4.23 &  0.14  & $-$33.9  &  1.41 \\

2011 Feb 05 & BM303p &  Core  &  194   &   19  &   Ref  &   0    &   Ref    &  0  \\*
            &        &   E    &   75   &    8  &   0.08 &   0.02 &    53.0  &  0.08  \\*
            &        &   D    &  21.9  &   2.3 &   0.38 &   0.02 & $-$25.2  &  0.30  \\*
            &        &   --   &   4.7  &   1.0 &   0.93 &   0.06 & $-$40.6  &  0.33  \\*
            &        &   C3*  &  11.2  &   1.9 &   1.52 &   0.08 & $-$36.6  &  0.59  \\*
            &        &   --   &   5.0  &    4  &   3.1  &   0.5  & $-$30.8  &  1.59  \\
         
2011 Feb 25 & BG207b &  Core  &  203   &  20   &   Ref  &  0     &   Ref    &  0 \\
            &        &   E    &  224   &  22   &   0.04 &  0.04  &    24.6  &  0.14 \\
            &        &   D    &   29   &   3   &   0.31 &  0.04  &  $-$5.2  &  0.33  \\
            &        &   --   &   10.5 &   1.1 &   0.69 &  0.04  & $-$42.3  &  0.38 \\
            &        &   C3   &   14.4 &   1.5 &   1.30 &  0.07  & $-$35.4  &  0.77  \\
            &        &   --   &    2.0 &   0.4 &   2.78 &  0.13  & $-$46.7  &  0.48  \\
            &        &   C1   &    5.6 &   0.8 &   4.01 &  0.14  & $-$33.1  &  0.71  \\
         
2011 Mar 01 & BM303q &  Core  &  168   &   17  &   Ref  &   0    &   Ref    &  0  \\*
            &        &   E    &  131   &   13  &   0.06 &   0.03 & $-$76.3  &  0.07  \\*
            &        &   D    &  20.2  &   2.2 &   0.36 &   0.03 & $-$8.1   &  0.17  \\*
            &        &   --   &  10.0  &   1.3 &   0.53 &   0.03 & $-$52.4  &  0.30  \\*
            &        &   C2   &  10.8  &   2.8 &   1.78 &   0.21 & $-$32.6  &  0.87  \\*
            &        &   --   &  14.6  &    3  &   3.54 &   0.17 & $-$26.0  &  0.91  \\

2011 Mar 29 & BG207c &  Core  &  189   &  19   &   Ref  &  0     &   Ref    &  0  \\
            &        &   E    &  170   &  17   &   0.06 &  0.03  &    37.8  &  0.11  \\
            &        &   D    &  27.6  &  2.8  &   0.37 &  0.03  & $-$18.6  &  0.32  \\*
            &        &   --   &   6.3  &  0.7  &   0.78 &  0.03  & $-$35.3  &  0.29  \\*
            &        &   C3   &  12.6  &  1.4  &   1.29 &  0.05  & $-$34.5  &  0.56  \\*
            &        &   C1   &  11.8  &  1.6  &   3.98 &  0.11  & $-$34.9  &  1.20  \\ 

2011 Apr 22 & BM303r &  Core  &  127   &   13  &   Ref  &   0    &   Ref    &  0  \\*
            &        &   E    &  199   &   20  &   0.06 &   0.02 &    28.2  &  0.12  \\*
            &        &   D    &  33.2  &    3  &   0.36 &   0.02 & $-$20.8  &  0.33  \\*
            &        &   C3   &  15.1  &   2.0 &   1.09 &   0.15 & $-$33.9  &  0.61  \\*
            &        &   C2   &   3.0  &   0.6 &   1.93 &   0.02 & $-$32.4  &  0.13  \\*
            &        &   --   &  11.9  &    3  &   3.9  &   0.4  & $-$36.3  &  1.51  \\

2011 Apr 25 & BG207d &  Core  &  153   &  15   &   Ref  &   0    &   Ref    &  0  \\*
            &        &   E    &   97   &  10   &   0.06 &  0.03  &    88.9  &  0.06  \\*
            &        &   --   &   33   &   3   &   0.13 &  0.03  &    10.6  &  0.13  \\*
            &        &   D    &  17.3  &  1.8  &   0.38 &  0.03  & $-$27.4  &  0.28  \\*
            &        &   --   &   0.8  &  0.8  &   0.80 &  0.07  & $-$49.6  &  0.47  \\*
            &        &   C3   &  11.4  &  1.4  &   1.43 &  0.17  & $-$31.7  &  0.76  \\*
            &        &   C1   &   5.2  &  0.9  &   4.20 &  0.11  & $-$34.4  &  0.79  \\

2011 May 22 & BM303s &  Core  &  216   &   22  &   Ref  &   0    &   Ref    &  0  \\*
            &        &   E    &  148   &   15  &   0.08 &   0.02 &    50.5  &  0.08  \\*
            &        &   D    &  38.3  &    4  &   0.40 &   0.02 & $-$14.2  &  0.27  \\*
            &        &   C3   &  12.7  &   1.9 &   1.20 &   0.08 & $-$36.4  &  0.41  \\*
            &        &   C2   &  15.2  &    4  &   2.5  &   0.3  & $-$30.2  &  1.17  \\*
            &        &   --   &  11.2  &    5  &   4.7  &   0.7  & $-$32.5  &  1.48  \\

2011 May 31 & BG207e &  Core  &  116   &  12   &   Ref  &   0    &   Ref    &  0  \\*
            &        &   E    &  106   &  11   &   0.06 &   0.03 &    20.9  &  0.12  \\*
            &        &   D    &   15.9 &   1.6 &   0.45 &   0.03 & $-$18.2  &  0.26  \\*
            &        &   C3   &    7.6 &   0.9 &   1.24 &   0.05 & $-$33.8  &  0.44  \\*
            &        &   C2   &    1.9 &   0.5 &   2.34 &   0.10 & $-$37.1  &  0.46  \\*
            &        &   C1   &    5.9 &   1.4 &   4.18 &   0.24 & $-$36.1  &  1.23  \\

2011 Jun 12 & BM303t &  Core  &  114   &   11  &   Ref  &   0    &   Ref    &  0  \\*
            &        &   E    &  116   &   12  &   0.06 &   0.02 &    52.1  &  0.07  \\*
            &        &   --   &  32.1  &    3  &   0.29 &   0.02 & $-$14.1  &  0.25  \\*
            &        &   --   &  10.4  &   1.6 &   0.75 &   0.12 & $-$38.0  &  0.47  \\*
            &        &   C3   &   4.6  &   0.8 &   1.37 &   0.02 & $-$26.9  &  0.16  \\*
            &        &   C2   &  12.1  &    5  &   2.85 &   0.8  & $-$46.1  &  1.94  \\

2011 Jun 29 & BG207f &  Core  &  92    &   9   &   Ref  &   0    &   Ref    &  0  \\*
            &        &   E    &  84    &   8   &   0.07 &   0.02 &    11.7  &  0.12  \\*
            &        &   D    &  15.5  &   1.6 &   0.41 &   0.02 & $-$20.4  &  0.23  \\*
            &        &   --   &   1.9  &   0.3 &   0.63 &   0.02 & $-$37.3  &  0.00  \\*
            &        &   C3   &  10.6  &   1.3 &   1.26 &   0.13 & $-$35.5  &  0.64  \\*
            &        &   C1   &   3.9  &   1.4 &   4.10 &   0.4  & $-$34.9  &  1.19  \\

2011 Jul 21 & BM303u &  Core  &  117   &   12  &   Ref  &   0    &   Ref    &  0  \\*
            &        &   E    &   54   &   5   &   0.09 &   0.03 &    26.1  &  0.08  \\*
            &        &   D    &  21.7  &   2.4 &   0.41 &   0.03 & $-$18.5  &  0.22  \\*
            &        &   C3   &   9.7  &   2.3 &   1.26 &   0.12 & $-$30.7  &  0.55  \\*
            &        &   --   &   4.6  &   1.1 &   2.70 &   0.03 & $-$30.0  &  0.14  \\*
            &        &   --   &  18.8  &   5   &   3.97 &   0.27 & $-$33.6  &  1.17  \\

2011 Jul 28 & BG207g &  Core  &  118   &  12   &   Ref  &   0    &   Ref    &  0  \\*
            &        &   E    &   74   &   7   &   0.07 &   0.03 &  $-$6.1  &  0.17  \\*
            &        &   D    &  18.0  &   1.8 &   0.48 &   0.03 & $-$24.1  &  0.28  \\*
            &        &   C3   &   8.9  &   1.0 &   1.19 &   0.04 & $-$33.5  &  0.54  \\*
            &        &   C2   &   3.9  &   0.8 &   2.19 &   0.16 & $-$26.0  &  0.98  \\*
            &        &   C1   &   5.3  &   1.0 &   3.62 &   0.22 & $-$33.0  &  1.33  \\

2011 Aug 23 & BM303v &  Core  &  139   &   14  &   Ref  &   0    &   Ref    &  0  \\*
            &        &   E    &   76   &    8  &   0.07 &   0.04 & $-$11.7  &  0.17  \\*
            &        &   D    &  25.5  &   2.6 &   0.46 &   0.04 & $-$20.6  &  0.43  \\*
            &        &   C3   &  12.4  &   1.4 &   1.27 &   0.06 & $-$34.4  &  0.59  \\*
            &        &   C2   &   2.9  &   0.5 &   2.46 &   0.04 & $-$58.1  &  0.15  \\*
            &        &   --   &  12.4  &   2.1 &   4.48 &   0.19 & $-$32.3  &  1.44  \\

2011 Aug 29 & BG207h &  Core  &  129   &  13   &   Ref  &   0    &   Ref    &  0  \\*
            &        &   E    &   72   &   7   &   0.04 &   0.03 &    28.4  &  0.11  \\*
            &        &   --   &  18.5  &   1.9 &   0.29 &   0.03 &  $-$0.9  &  0.15  \\*
            &        &   D    &  14.3  &   1.5 &   0.50 &   0.03 & $-$27.1  &  0.33  \\*
            &        &   C3   &   8.7  &   1.0 &   1.23 &   0.07 & $-$31.4  &  0.56  \\*
            &        &   C1   &   6.5  &   1.2 &   4.32 &   0.19 & $-$36.8  &  1.21  \\

2011 Sep 16 & BM303w &  Core  &   77   &   8   &   Ref  &   0    &   Ref    &  0  \\*
            &        &   E    &   69   &   7   &   0.09 &   0.02 & $-$41.0  &  0.08  \\*
            &        &   --   &  26.4  &   3   &   0.31 &   0.09 & $-$23.6  &  0.53  \\*
            &        &   C3   &   7.3  &   2.1 &   1.25 &   0.16 & $-$55.2  &  0.57  \\*
            &        &   C2   &   6.5  &   1.2 &   2.16 &   0.05 & $-$45.6  &  0.27  \\*
            &        &   --   &   4.6  &   1.0 &   2.78 &   0.02 & $-$57.6  &  0.00  \\

2011 Sep 29 & BG207i &  Core  &  124   &  12   &   Ref  &   0    &   Ref    &  0  \\*
            &        &   E    &   98   &   10  &   0.05 &   0.04 &  $-$4.4  &  0.12  \\*
            &        &   --   &  25.1  &   2.5 &   0.39 &   0.04 & $-$19.8  &  0.26  \\*
            &        &   --   &   8.5  &   0.9 &   0.88 &   0.04 & $-$30.8  &  0.36  \\*
            &        &   C3   &   5.3  &   0.6 &   1.43 &   0.04 & $-$33.8  &  0.35  \\*
            &        &   C1   &  11.5  &   1.6 &   4.49 &   0.16 & $-$34.8  &  1.56  \\

2011 Oct 16 & BM353a &  Core  &  150   &   15  &   Ref  &   0    &   Ref    &  0  \\*
            &        &   E    &   79   &    8  &   0.05 &   0.03 &    31.5  &  0.14  \\*
            &        &   --   &  36.2  &    4  &   0.31 &   0.03 & $-$34.8  &  0.36  \\*
            &        &   --   &  16.1  &   2.0 &   0.86 &   0.06 & $-$34.5  &  0.46  \\*
            &        &   C2   &   4.6  &   0.8 &   1.99 &   0.03 & $-$48.4  &  0.14  \\*
            &        &   --   &   7.6  &    3  &   4.2  &   0.4  & $-$39.5  &  1.15  \\

2011 Oct 29 & BG207j &  Core  &  119   &  12   &   Ref  &   0    &   Ref    &  0  \\*
            &        &   E    &  102   &   10  &   0.03 &   0.03 &   173.9  &  0.11  \\*
            &        &   --   &  27.2  &   2.7 &   0.24 &   0.03 & $-$16.2  &  0.27  \\*
            &        &   --   &  11.6  &   1.2 &   0.66 &   0.03 & $-$32.5  &  0.35  \\*
            &        &   C3   &   9.6  &   1.0 &   1.25 &   0.04 & $-$30.7  &  0.46  \\*
            &        &   C1   &  10.9  &   1.7 &   4.26 &   0.18 & $-$34.4  &  1.56  \\

2011 Nov 28 & BG207k &  Core  &  156   &  16   &   Ref  &   0    &   Ref    &  0  \\*
            &        &   C4   &  106   &   11  &   0.05 &   0.03 &   167.9  &  0.08  \\*
            &        &   --   &   30   &   3   &   0.17 &   0.03 &    11.1  &  0.12  \\*
            &        &   --   &  18.8  &   1.9 &   0.40 &   0.03 & $-$22.9  &  0.33  \\*
            &        &   --   &   5.7  &   0.6 &   0.91 &   0.03 & $-$38.5  &  0.25  \\*
            &        &   C3   &   7.9  &   0.9 &   1.39 &   0.05 & $-$32.2  &  0.50  \\*
            &        &   C1   &   9.2  &   1.3 &   3.81 &   0.13 & $-$36.1  &  1.26  \\

2011 Dec 02 & BM353b &  Core  &   92   &   9   &   Ref  &   0    &   Ref    &  0  \\*
            &        &   --   &   69   &   7   &   0.07 &   0.02 & $-$29.0  &  0.05  \\*
            &        &   E    &  42.0  &   4   &   0.07 &   0.02 &    51.6  &  0.13  \\*
            &        &   --   &  17.4  &   2.0 &   0.41 &   0.02 & $-$19.0  &  0.31  \\*
            &        &   C3   &  16.0  &   2.5 &   1.14 &   0.08 & $-$57.2  &  0.65  \\*
            &        &   --   &   4.7  &   0.9 &   5.10 &   0.02 & $-$39.2  &  0.15  \\

2011 Dec 23 & BG207l &  Core  &  150   &  15   &   Ref  &   0    &   Ref    &  0  \\*
            &        &   E    &  143   &   14  &   0.05 &   0.02 &    32.8  &  0.13  \\*
            &        &   --   &  19.2  &   2.0 &   0.44 &   0.02 & $-$20.8  &  0.29  \\*
            &        &   --   &   8.6  &   1.1 &   0.87 &   0.09 & $-$37.1  &  0.45  \\*
            &        &   C3   &   6.6  &   1.0 &   1.28 &   0.12 & $-$28.3  &  0.45  \\*
            &        &   C1   &   5.0  &   1.1 &   4.37 &   0.12 & $-$34.2  &  0.59  \\

\end{longtable}
}% End \longtab
\twocolumn
\end{document}